\documentclass[aps,prd,superscriptaddress,nofootinbib,amsmath,amsfonts,preprintnumbers,groupedaddress,showpacs,10pt,english]{revtex4-1}
\usepackage{amsmath}
\usepackage{amssymb}
\usepackage{babel}
\usepackage{wrapfig}
\usepackage{cancel}

\usepackage{relsize,exscale}
\makeatletter

\usepackage{array,multirow,graphicx}
\usepackage{dcolumn}
\usepackage{newlfont}
\usepackage{bm}
\usepackage[colorlinks,citecolor=blue,urlcolor=blue,linkcolor=blue]{hyperref}
\usepackage[figtopcap]{subfigure}
\usepackage{color}

\newcommand{\pont}{{\,^\ast\!}R\,R}

\newcommand\be{\begin{equation}}
\newcommand\ba{\begin{eqnarray}}
\newcommand\ee{\end{equation}}
\newcommand\ea{\end{eqnarray}}

\begin{document}

\date{\today}
\title{Solutions of a slowly rotating Kerr flat-horizon black hole in dynamical Chern-Simons modified gravity}

\author{G.G.L. Nashed}
\email{nashed@bue.edu.eg}
\affiliation {Centre for Theoretical Physics, The British University, P.O. Box
43, El Sherouk City, Cairo 11837, Egypt}
\author{Kazuharu Bamba}%
\email{bamba@sss.fukushima-u.ac.jp}
\affiliation{Faculty of Symbiotic Systems Science,
Fukushima University, Fukushima 960-1296, Japan}
\begin{abstract}
Solutions pertaining to a Kerr black hole with a flat horizon undergoing gradual rotation are explored in the context of gravitational theories modified by dynamical Chern-Simons terms with cylindrical metrics, which approach asymptotically the anti de Sitter spacetime. It is shown that the cross-term of a metric component is unaffected by the perturbations of the Chern-Simons scalar independently of whether the dynamical Chern-Simons field equation is uncharged or charged with an electric field. From this result, it is ensured that the Chern-Simons scalar field can affect the spaces of the metric that approach asymptotically the flat spacetime only.
\end{abstract}





\maketitle

\section{Introduction}
\label{intro}
Modified gravity theories (as well as dark energy models) have been studied to account for the late-time cosmic acceleration (including inflation in the early universe) and solve cosmological and astrophysical issues which cannot be explained in the context  of General Relativity (GR) \cite{padmanabhan2007, Durrer2008, Sotiriou:2008rp, DeFelice2010, Nojiri:2010wj, Capozziello2011, Clifton:2011jh, Joyce:2014kja, Cai:2015emx, Nojiri:2017ncd}.

When we consider the physics of black holes, however, it is not a trivial task to choose one theory among different  modified gravity theories. It is well-known that the solutions of the Schwarzschild and Kerr black holes are popular in different alternative constructions, and that the dynamical property can be analyzed in the regime of strong gravitational field for the observational limits. Hence, it is more useful to discuss differences from GR with an appropriate class of parameters, by measuring the shift starting from established metric solutions and ensuring that the Schwarzschild and Kerr geometries can be smoothly derived under certain suitable conditions. Such an idea has been explored in, e.g., Ref. \cite{Johannsen:2011dh}, where alterations to the Kerr solution occur organically, accommodating arbitrary spin values, in the context of modified gravity, specifically in scenarios characterized by rotating solutions, it pertains to a stationary, axisymmetric, and asymptotically flat metric, irrespective of the specific theoretical framework in use.

This approach offers the advantage of addressing the majority of issues that typically surface in perturbation theory within GR. These challenges primarily stem from the broken  of the no-hair theorems \cite{Gair:2007kr,Johannsen:2010xs,Sotiriou:2015pka,Herdeiro:2014goa,Herdeiro:2015waa,Cardoso:2016ryw}, which significantly limit the forecasts of certain  scenarios, such as maximum mass-disparity inspirals \cite{Barack:2006pq} alternatively, the orbital dynamics of stars and pulsars in proximity to black holes \cite{Wex:1998wt,Will:2007pp,Broderick:2013rlq}. Nonetheless, the absence of a clearly established model elucidating the genesis of the metric presented in Ref.~\cite{Johannsen:2011dh} constrains its utility in scenarios extending beyond black holes. This raises questions about how such a solution can be systematically derived..

Due to the above mentioned  point, the Chern-Simons (CS) modified gravity is a beneficial theory. In the following, we first explain the significance of the CS theories as well as the previous related studies.

The CS theory could address the issues of black holes and quantum aspects of gravitation in a single theoretical framework. The CS gravitational theory as influenced by the alteration of  electrodynamics, as presented by Jackiw and Pi in their work \cite{Jackiw:2003pm,Carroll:1989vb}. The Lagrangian for the Maxwell field in the theory of $U(1)$ gauge undergoes modification through the inclusion of a (pseudo)-scalar field $\vartheta(x)$ that is linked to the topology of the Pontryagin density of $U(1)$  $^{*}FF\equiv ^{*}F_{\mu\nu}F^{\mu\nu}$. Despite keeping the gauge invariance, such modification admits for the violation of the Lorentz/CPT symmetry \cite{Carroll:1989vb}. This is understood from the expression of the Carroll-Field-Jackiw form ${\mathit v}_{\mu}\,^{*}F^{\mu\nu}A_{\nu}$, with ${\mathit v}_{\mu}\equiv \partial_{\mu}\vartheta$ being  the axial vector which is the source of the violation of the Lorentz/CPT symmetry. If we extend the theoretical construction, ${\mathit v}_{\mu}$ connects to one of the coefficients associated with Lorentz CPT violation within the framework of the extension to the standard model \cite{Colladay:1996iz, Colladay:1998fq,Kostelecky:2003fs}. In the context of CS modified gravity, as seen in modifications like the CS alteration to Maxwell's electrodynamics, where an introduction of a non-minimal coupling is made between $\vartheta(x)$ and the gravitational Pontryagin density  as follows $^{*}RR \equiv {}^{*}R^{\mu\nu\alpha\beta}R_{\nu\mu\alpha\beta}$ (denoted in the Lagrangian of GR).

The relation between the (anti) de-Sitter (A)dS and the conformal field theory (CFT) \cite{Maldacena:1997re,Witten:1998qj,Witten:1998zw,Gubser:1998bc,Aharony:1999ti} has gained a lot of interest in the asymptotic AdS spacetimes and their corresponding thermodynamics. A number of phenomena in gauge field theories such as phase transitions, the encoding of confinement, and the conformal anomalies in the physics of semi-classical asymptotically AdS black holes has been investigated \cite{Witten:1998qj,Witten:1998zw,Henningson:1998gx}.  Aside from its critical role in the preceding duality, asymptotic (A)dS has different attractive characteristics. In contrast to the asymptotic flat situation where $\Lambda = 0$, the case with a negative cosmological constant ($\Lambda < 0$) can be permitted for different kinds of topologies for the black holes horizons. Such horizons could have different forms i.e.,  hyperbolic, spherical or planar, global identifications can result in the formation of cylinders (in the planar case) or the tori and Riemann surfaces (in the hyperbolic case) in four dimensions \cite{Mann:1997iz}. In the present paper, we apply a rotating flat horizon spacetime to the CS dynamical theory and try to derive a black hole solution.

The structure of the current investigation is outlined as follows. In Sec. \ref{ABC}, we present the cornerstone of the CS gravitational theory. In Sec. \ref{axisym}, we explain the necessary tools that we use in the techniques of perturbations. In Sec. \ref{slow-rot1}, we give a brief review of the flat horizon black hole (BH) solution.   In Sec. \ref{slow-rot}, we apply the metric or rotating flat horizon to the dynamical CS field equations and show that the scalar field does not affect the metric component, $g_{t,\theta}$. In Sec. \ref{charg}, we take the procedure described in Sec. \ref{slow-rot} with a charge rotating flat horizon, and we demonstrate that the scalar field has no effect the metric component $g_{t\,\theta}$.In conclusion, we conclude our investigation by presenting the main findings in Sec. \ref{conclusions}.

The predominant choice for spacetime signature is mostly positive, and the gravitational coupling is defined as $\kappa^2=8\pi$, describing in  geometric units where $G=c=1$.   Levi-Civita tensor which is denoted as $\varepsilon_{\mu\nu\rho\sigma}$, can be expressed in terms of fully antisymmetric symbol $\varepsilon_{\mu\nu\rho\sigma}$, where $\varepsilon_{0123}=1$.

\section{Chern-Simons amended gravitational theory}
\label{ABC}
In this section we are going to present the main tools of the CS gravitational theory that support the present study \cite{Nashed:2023cyr}.

In CS gravitational theory, the Lagrangian density takes the form:
\be
\label{CSaction}
L =L_{1} + L_{2} +  L_{3} + L_{4}\,.
\ee
Here,
$L_1$ is the Einstein-Hilbert Lagrangian involving the cosmological constant, defined as:
\be
\label{EH-action}
L_{1} = \kappa \int_{V} d^4x  \sqrt{-g} ( R-2\Lambda),
\ee
$L_{2}$ is the Lagrangian of the scalar field, given by:
\be
\label{Theta-action}
L_{2} = - \frac{\alpha_1 }{2} \int_{V} d^{4}x \sqrt{-g} \left[ g^{\alpha \beta}
\left(\nabla_{\alpha} \vartheta\right) \left(\nabla_{\beta} \vartheta\right) + 2 V(\vartheta) \right], \quad
\ee
${L}_{3}$ is the Chern-Simons Lagrangian, expressed as:
\be
\label{CS-action}
L_{3}= \frac{\alpha}{4} \int_{V} d^4x  \sqrt{-g} \;
\vartheta \; \pont\,,
\ee
$L_{4}$ is the Lagrangian of the matter field, represented as:
\be\label{MM}
L_{M}= \int_{V} d^{4}x \sqrt{-g} {L}_{mat.}.
\ee

 The following symbols are used through this study: $\alpha$ and $\alpha_1$ are {dimensional} constants; $\nabla_{\beta}$ is the covariant derivative;
 $g$ is the  metric determinant; $R$ is the Ricci scalar tensor, $\Lambda$ is the cosmological constant that can rewritten as $\Lambda=\frac{-3}{\iota^2}$, and $\kappa = \frac{1}{16 \pi G}$. The symbol $\pont$ refers to the Pontryagin density and is defined as
\be
\label{pontryagindef}
\pont= {\,^\ast\!}R^\alpha{}_{\beta}{}^{ \gamma \delta} R^{\beta}{}_{\alpha \gamma \delta}\,,
\ee
with ${\,^\ast\!}R^\alpha{}_{\beta}{}^{ \gamma \delta}$ the dual Riemann-tensor, defined as:
\be
\label{Rdual}
{\,^\ast\!}R^\alpha{}_{\beta}{}^{ \gamma \delta}=\frac12 \varepsilon^{ \gamma \delta \gamma_1 \delta_1}R^\alpha{}_{\beta \gamma_1 \delta_1}\,,
\ee
where $\varepsilon^{ \gamma \delta \gamma_1 \delta_1}$ is  completely  anti-symmetric tensor and  $\varepsilon^{0123} = -1$.

The expression for the CS scalar field $\vartheta$ is the one that parameterizes the deformation in GR. If $\vartheta = const.$, GR is recovered since the Pontryagin density becomes a total derivative of the CS topological field which is defined as:
\be
\nabla_\alpha \Psi^a = \frac{1}{2} \pont \,,
\label{eq:curr1}
\ee
where
\be
\Psi^\alpha =\varepsilon^{\alpha \beta \gamma \delta} \Gamma^{\rho}_{{\beta} \epsilon} \left(\partial_{\gamma}\Gamma^{\epsilon}_{\delta \rho}+\frac{2}{3} \Gamma^{\epsilon}_{\gamma \gamma_1}\Gamma^{\gamma_1}_{\delta \rho}\right)\,,
\label{eq:curr2}
\ee
with $\Gamma$ the second kind of Christoffel symbol.
By using Eq. \eqref{eq:curr2}, we describe $L_{3}$ in the form \cite{Yunes:2007ss}
\be
\label{CS-action-K}
L_{{CS}} =  \alpha
\left( \vartheta \; {\Psi}^{\alpha} \right)|_{\partial {{V}}}
-
 \frac{\alpha}{2} \int_{{{V}}} d^4x  \sqrt{-g} \;
\left(\nabla_{\alpha} \vartheta \right) {\Psi}^{\alpha}\,.
\ee
In general, we do not take into account the first term in Eq. \eqref{CS-action-K} because it is described at the boundary of the spacetime~\cite{Grumiller:2008ie}. On the other hand, the second term is regarded as the CS correction term.

The variation of the Lagrangian \eqref{CSaction} with respect to (w.r.t) the metric, and the CS coupling leads to the following field equations:
\ba
\label{eom}
&&R_{ab} -2g_{\alpha \beta} \Lambda+ \frac{\alpha_1}{\kappa} C_{\alpha \beta} = \frac{1}{2 \kappa} \left(T_{\alpha \beta} - \frac{1}{2} g_{\alpha \beta} [T -4\Lambda] \right),
\\
\label{eq:constraint}
&&\alpha_1 \; \square \vartheta = \alpha_1 \; \frac{dV}{d\vartheta} - \frac{\alpha}{4} \pont\,.
\ea
Here the Ricci second order tensor is $R_{ab}$ while $\square = \nabla_{\mu} \nabla^\mu$ is the D'Alembertian operator.
The term  $C_{ab}$ is the C-tensor, defined as
\be
\label{Ctensor}
C^{\alpha \beta} = v_\gamma
\varepsilon^{\gamma \gamma_1\gamma_2(\alpha}\nabla_{\gamma_2}R^{\beta){}_{\gamma_1}}+v_{\alpha_1\alpha_2}{\,^\ast\!}R^{\alpha_2(\alpha \beta)\gamma}\,,
\ee
where
\be
\label{v}
v_\alpha=\nabla_\alpha\vartheta\,,\qquad
v_{\alpha \beta}=\nabla_\alpha\nabla_{\beta}\vartheta\,.
\ee
The term $T_{\alpha \beta}$ is the total stress-energy tensor which can be expressed as: 
\be\label{Tab-total}
T_{\alpha \beta} = T^{{mat}}_{\alpha \beta} + T_{\alpha \beta}^{\varrho}\,.
\ee
Here, $T^{mat}_{\alpha \beta}$ is the matter source, which we do not consider in this study, and $T_{\alpha \beta}^{\vartheta}$ is defined as
\ba
\label{Tab-theta}
&&T_{\alpha \beta}^{\vartheta}
=   \alpha_1  \left[  \left(\nabla_{\alpha} \vartheta\right) \left(\nabla_{\beta} \vartheta\right)
    - \frac{1}{2}  g_{\alpha \beta}\left(\nabla^{\alpha_1} \vartheta\right) \left(\nabla^{\alpha_1} \vartheta\right)
-  g_{\alpha \beta}  V(\vartheta)  \right]\,.
\ea
The principle of strong equivalence which is denoted as ($\nabla_{\beta} T^{\alpha \beta}_{{mat}} = 0$), holds true in the case of CS theory when the field equations \eqref{eq:constraint} for the scalar field $\vartheta$ are fulfilled. This is due to the fact that the derivative of Eq.~\eqref{eom}, the first term,   becomes nullified which is the characteristics of the Bianchi identities. However, the next term  which is linked to the Pontryagin density in a proportionate manner, as per the following expression:
\be
\label{nablaC}
\nabla_\alpha C^{\alpha \beta} = - \frac{1}{8} v^\beta \pont.
\ee
The confirmation of Eq. \eqref{nablaC} leads  to  Eq.~\eqref{eq:constraint}.

The dimensional characteristics of the constants employed in this investigation and the scalar field $\vartheta$ are considered as follows.
The fixation of one of
$(\alpha,\alpha_1,\vartheta)$ fixes the units of the others.
For example, if the unit of the CS scalar field is $[\vartheta] = l^{b}$,
$[\alpha] = l^{2 -b}$ and $[\alpha_1] = l^{-2b}$, with $l$ being the standard unit of length. To make the CS scalar $\vartheta$ a dimensionless quantity, $[\alpha] = l^{2}$ and $\alpha_1$ must be dimensionless in the scalar-tensor theories as usual~\footnote{We are going to put $G = c = 1$, and thus, the Lagrangian  takes the measurement of $l^{2}$. Therefore, when employing natural units then $G= c = 1$, therefore, the Lagrangian will have no dimensions, and in the event that $[\vartheta] = l^{b}$ then $[\alpha] = l^{-b}$ and $[\alpha_1] = l^{-2 b - 2}$.}. Another option is $\alpha = \alpha_1$, which results in $[\vartheta] = l^{-2}$ if $L_2$ and the Lagrangian of CS are on equal footing. Since no formulation forces us to choice special measurement units for $\vartheta$, we keep it unspecified, as a result of previous studies that have opted for a different selection.

\section{Rotating flat horizon BH solution}\label{axisym}

In this section, we use two different schemes  and resolve the modified dynamical CS equation of motions till the second order perturbation term to examine the properties of the rotating flat horizon BH solution in the dynamical CS theory of gravity.

\subsection{The approximate method}
\label{approx}

The procedures of two approximations developed in this work are as follows.
$a$ is the slow rotation parameter and we consider the  approximations of small-coupling. In this process, one considers the amended CS like  a minor alteration of the  GR, enabling us to make an assumption about the metric decomposition (up to the second order) in the following manner:
\ba\label{small-cou-exp0}
&&g_{ab} = g_{ab}^{(0)} + \xi g^{(1)}_{ab}(\vartheta) + \xi^{2} g^{(2)}_{ab}(\vartheta)\,, \nonumber\\
%
\ea
where $g_{ab}^{(0)}$ represents the underlying metric which verifies the  equation of motions of GR, like (A)dS-Kerr metric, while $g_{ab}^{(1)}(\vartheta)$ and $g_{ab}^{(2)}(\vartheta)$ constitute the perturbations at both first and second orders of the CS theory. The parameter $\xi$ represents the degree of approximation within the small-coupling scheme.

Moreover, We can re-express the base configuration and the $\xi$-perturbations expressed in powers of  (A)dS-Kerr through the utilization of the slow rotation approximation. Thus, the metric background and the metric perturbations give:
\ba
\label{small-cou-exp}
g_{ab}^{(0)} &=& \eta_{ab}^{(0,0)} + \epsilon \; h_{ab}^{(1,0)} + \epsilon^{2} h_{ab}^{(2,0)},
\nonumber \\
\xi g_{ab}^{(1)} &=& \xi h_{ab}^{(0,1)} + \xi \epsilon \; h_{ab}^{(1,1)} + \xi \epsilon^{2} h_{ab}^{(2,1)},
\nonumber \\
\xi^{2} g_{ab}^{(2)} &=& \xi^{2} h_{ab}^{(0,2)} + \xi^{2} \epsilon \; h_{ab}^{(1,2)} + \xi^{2} \epsilon^{2} h_{ab}^{(2,2)}\,,
\ea
where $\epsilon$ represents the degree of expansion within the slow rotation approximation.
It is noted that the notation $h^{o\;p}_{a\;b}$ is the expressions of ${\cal{O}}(o\;p)$, which points to an expression of ${\cal{O}}(\epsilon^{o})$ and ${\cal{O}}(\xi^{p})$. In Eq.~\eqref{small-cou-exp}, $\eta_{ab}^{(0,0)}$ represents the base metric in the absence of the rotation parameter, i.e., $a= 0$, whereas $h_{ab}^{(1,0)}$, $h_{ab}^{(2,0)}$, $\xi_{a}^{(1,0)}$ and $\xi_{a}^{(2,0)}$ constitute the perturbations of the base metric at both the first and second orders.
We obtain the bivariate expansion by using the two independent parameter $\xi$ and $\epsilon$,which take on the following metric form with perturbations up to the second order:
\ba
&&g_{ab} = \eta_{ab}^{(0,0)} + \epsilon h_{ab}^{(1,0)} + \xi h_{ab}^{(0,1)} + \epsilon \xi h_{ab}^{(1,1)} + \epsilon^{2} h_{ab}^{(2,0)} + \xi^{2} h_{ab}^{(0,2)}\,.\nonumber\\
\ea
The terms  ${\cal{O}}(1,0)$ or ${\cal{O}}(0,1)$ corresponding to the first-order while ${\cal{O}}(2,0)$, or ${\cal{O}}(0,2)$ or ${\cal{O}}(1,1)$  corresponding to the second-order.

Slow-rotation involves expanding the parameter  $a$, and such expansion must be dimensionless, represented as $a/M$. Hence, any equation multiplied by $\epsilon^n$ is  ${\cal{O}}\left((a/M)^n\right)$. Before applying the above procedure to (A)dS-Kerr solution, we present a brief review of this BH in the next section.

\section{Flat horizon black hole solution}\label{slow-rot1}
The action for GR with the Maxwell field of asymptotic AdS spacetime  is given by

\begin{eqnarray}
L= -{1 \over 16 \pi G_d}\int_{\cal M} d^{4}x \sqrt{-g}\left(R+{6
\over \iota^2}\right)\,.
\end{eqnarray}
The variation of the action w.r.t. the metric gives the following
field equations:
\be
G_{\mu\nu}-\Lambda g_{\mu\nu}=2\,T_{\mu\nu}\,.
\ee

Lemos obtained a four-dimensional   solution having rotation and  flat horizon, is presented in \cite{Lemos:1994xp, Lemos:1995cm}, and it is expressed in the following manner:
\be \label{sol}
ds^2=-Q(r)\,(\Xi\,dt-a\,d\theta)^2+{dr^2 \over Q(r)}+{r^2 \over
\iota^4}(a\,dt-\Xi \iota\,d\theta)^2+{r^2\over \iota^2} \,d{\rho}^2\,,
\ee
where $Q(r)$ is given by
\be
Q(r)={r^2 \over \iota^2}-{M \over r}
.\\
\ee
The entities $\theta$ and $\rho$ are supposed  to be $0\leq \theta
< 2\pi $ and $-\infty<\rho<\infty$, respectively. In this context, $a$ represents the   rotation, $M$ is the   mass, and $\Xi$ is figures as $\Xi=\sqrt{1+a^2/\iota^2}$.
The solution given in \eqref{sol} is investigated in previous works like the one presented in \cite{Lemos:1994xp, Lemos:1995cm, Dehghani:2002rr, Cardoso:2001vs, Awad:2002cz}, and preserves supersymmetry \cite{Lemos:2000wp}. It's evident that a  transformation exists, which linking the   solution with $a$ to the  one without $a$, and this transformation is defined as:
\be t'=\Xi\,t-a\,\theta
\hspace{1.0 in} \theta'={a \over l^2}\,t-\Xi\,\theta \,.
\ee
Due to its combination of the compactified $\theta$ coordinate and the time coordinate $t$. The above transformation is locally permissible however, it is not valid globally~\cite{Lemos:1994xp}. References \cite{Lemos:1994xp, Stachel:1981fg}have demonstrated that in cases where the first Betti number of the manifold is non-zero, there exist no global diffeomorphisms capable of mapping the two metrics \cite{Lemos:1994xp,Stachel:1981fg}. References \cite{Lemos:1994xp, Stachel:1981fg} have demonstrated that solution \eqref{sol} possesses a first Betti number of one when cylindrical or toroidal horizons are present.

\section{Slowly rotating (A)dS BH solution}\label{slow-rot}

In the regime of slow rotation, the expansion of the background metric can be characterized by the following line element:
\ba
\label{slow-rot-ds2}
&&ds^{2} = -Q  \left[1 + Q_1(r)\right] dt^{2}
+ \frac{1}{Q}  \left[1 + Q_2(r)\right] dr^{2}
+\frac{r^{2}}{\iota^2}\left[1 + Q_3(r) \right] \left[ d\phi  - \omega(r) dt \right]^{2}+ \frac{r^{2}}{\iota^2} \left[1 + Q_4(r) \right] d\rho^{2}\,,\nonumber\\
&&\ea
where $Q$ takes the form $Q = \frac{r^2}{\iota^2} - \frac{M}{r}$.
This is the (A)dS-Schwarzschild solution which we take as the background metric, and $M$ is the mass of the  BH in the absence of the CS expression.
In Eq.~\eqref{slow-rot-ds2}, we use the cylindrical coordinates  with the perturbations of the metric are $Q_1(r)$, $Q_2(r)$, $Q_3(r)$, $Q_4(r)$ and $\omega(r)$.

The metric~\eqref{slow-rot-ds2} can be rewritten similarly to the one presented in Refs.~\cite{Lemos:1994xp,Lemos:1995cm}, and The expansions of the metric perturbations need to be carried out in asymptote of $\xi$ as well as $\epsilon$. The expression for the second-order expansion is given by:
\ba \label{cons}
 Q_1(r) &=& \epsilon \; Q_1{_{(1,0)}} + \epsilon\;  \xi \;  Q_1{_{(1,1)}} + \epsilon^{2} \;  Q_1{_{(2,0)}},
 \nonumber \\
 Q_2(r) &=& \epsilon \; Q_2{_{(1,0)}} + \epsilon\;  \xi \;  Q_2{_{(1,1)}} + \epsilon^{2} \; Q_2{_{(2,0)}},
 \nonumber \\
 Q_3(r) &=& \epsilon\; Q_3{_{(1,0)}} + \epsilon\;  \xi \;  Q_3{_{(1,1)}} + \epsilon^{2} \;  Q_3{_{(2,0)}},
 \nonumber \\
 Q_4(r) &=& \epsilon\; Q_4{_{(1,0)}} + \epsilon\;  \xi \;  Q_4{_{(1,1)}} + \epsilon^{2} \;  Q_4{_{(2,0)}},
 \nonumber \\
  \omega(r) &=& \epsilon\; \omega_{(1,0)} + \epsilon \;  \xi \; \omega_{(1,1)} + \epsilon^{2} \; \omega_{(2,0)}.
 \ea
Equations~\eqref{cons} have no terms of ${\cal{O}}(0,0)$ because it exists in the (A)dS-Schwarzschild structure of Eq.~\eqref{slow-rot-ds2}, in which we have bivariated the diagonal components of (A)dS-Kerr in terms of the rotation parameter $a$. Moreover, it is supposed that if the rotation parameter of (A)dS-Kerr is vanishing, we can show that the Schwarzschild-(A)dS  is a solution. This ensures that all the terms of ${\cal{O}}(0,s)$ are vanishing. Therefore, the CS expression  must be linear in terms of the (A)dS-Kerr rotation.
When the rotation parameter $a$ is perturbed,
the metric perturbation proportional to $\xi^{0}$ up to the first order takes the following form
\ba\label{0e}
Q_1{_{(1,0)}} &=& Q_2{_{(1,0)}} = Q_3{_{(1,0)}} = Q_4{_{(1,0)}} =0\,, \qquad
\omega_{(1,0)}=-2\frac{M}{r}\frac{a}{\iota}\,,
\ea
 and up to second order as
\ba
Q_1{_{(2,0)}} &=&\frac {M}{r}\frac{a^2}{\iota^2}\,,\qquad  \qquad Q_2{_{(2,0)}}=Q_4{_{(2,0)}}= \omega_{(2,0)} = 0\,,\qquad \qquad
Q_3{_{(2,0)}}= \frac{M}{r}\frac{a^2}{\iota^2}\,,
\ea
which corresponds to the (A)dS Schwarzschild solution when the rotation parameter vanishes, namely, $a=0$ \cite{Nashed:2021xtt,Nashed:2021sey,Nashed:2020kdb}.
All the fields are bivariate in terms of the rotation parameter $a$, in addition within the framework of the small-coupling approximation, incorporating the CS field.
To derive the leading-order of the scalar field  $\vartheta$,
we analyze the evolution equation in Eq.~\eqref{eq:constraint}.
This equation yields $\partial^{2} \vartheta \sim (\alpha_1/\alpha) \pont$, from which the Pontryagin density is vanishing  till order $a/M$.
As a result, the CS first order behavior should behaves as $\vartheta \sim (\alpha_1/\alpha) ((a/M))$, i.e.,  $\propto \epsilon$.
We should have $\vartheta^{(0,s)} = 0$ for all $s$ should the Schwarzschild-(A)dS metric which is the unique solution when $a$ is zero.
In the present paper, we describe  slowly  rotating solution.
By utilizing Eq. \eqref{eq:constraint} in Eq. \eqref{slow-rot-ds2} and with Eq. \eqref{cons}, we get
\be
\label{th-ansatz}
\vartheta = \epsilon \; \vartheta^{(1,0)}(r) + \epsilon \;  \xi \; \vartheta^{(1,1)}(r) + \epsilon^{2} \;\vartheta^{(2,0)}(r)\,.
\ee
By applying the method outlined above for solving the modified field equations and highlighting the evolutionary equation for the dynamical CS scalar field, up to   ${\cal{O}}(1,0)$ we acquire the form
\ba
\label{1st-eq1}
&&Q \vartheta^{(1,0)}_{,rr} + \frac{1}{r} \vartheta^{(1,0)}_{,r} \left(\frac{4r^2}{\iota^2} - \frac{M}{r} \right) =0 \,.
\ea
 The solution of the above differential equation \eqref{1st-eq1} is given by
\be\label{theta-sol-SR111}
\vartheta^{(1,0)}(r) =\frac{c_0}{3\,M\,\iota^2}\ln\left(\frac{1}{1-\frac{M}{r}\,\frac{\iota^2}{r^2}}\right).
\ee
where $c_0$ is a dimensional constant of integration.

The asymptotic form of Eq. \eqref{theta-sol-SR111} is expressed as
\ba
\label{theta-sol-SR111}
&&\vartheta^{(1,0)}(r)\approx \frac{c_0}{r^3}\left(1+\frac{M}{2r}\frac{\iota^2}{r^2}\right)+{\cal O}\left(\frac{1}{r^9}\right)\,.
\ea

Since we present the dynamical CS field, we derive the dynamical CS corrections of the metric perturbations. We omit the sharing of the energy-momentum tensor in Eq. \eqref{Tab-theta} to the metric perturbation because of its sharing to the modified equation of motions in Eq. \eqref{eom} up to ${\cal{O}}(2,1)$. The modified Einstein equations are divided into two sets.
The initial set constructs a  system of  differential equations
that have ${h_1}^{(1,1)}$, ${h_2}^{(1,1)}$, ${h_3}^{(1,1)}$ and ${h_4}^{(1,1)}$, that are the components $(t,t)$, $(r,r)$, $(r,\theta)$, $(\theta,\theta)$ and $(\rho,\rho)$. The second set gives one   equation of $\omega^{(1,1)}$, that is the $(t,\theta)$.

The initial group is not dependent on the dynamics of the CS, $\vartheta$, and therefore its contribution vanishes identically. Hence, we investigate the second set and consider $(t,\theta)$-component, which leads to
\ba
\label{V1}
Q\omega^{(1,1)}_{rr}-\frac{4Q}{r}\omega^{(1,1)}_r-\frac{6\alpha_1}{\iota^2}\omega^{(1,1)} =6\alpha_1\frac{M}{r}\frac{a}{\iota}\frac{\alpha_1}{r\iota}\,. \ea
The equation above is a non-homogeneous differential equation, possessing both homogeneous and particular solutions. The homogeneous solution is structured as follows:
 \be\label{part}
 \omega^{(1,1)}=r^\beta\left(1-\frac{M}{r}\frac{\iota^2}{r^2}\right)\left\{c_1H([b],c,d)+c_2H([-b],-c,d)\right\}\,,
\ee
where $H$ denotes a hypergeometric function\footnote{The hypergeometric function is defined as \[ H(n,d,z)=Sum(z^k/k! *product (pochhammer(n[i], k), i=1\cdots p)/product (pochhammer(d[j], k), j=1..p),k=0\cdots \infty)\,.\]}
and $\beta=\frac{-6+\sqrt{96\beta+36}}4$, $b=[\frac{6-\sqrt{96\beta+36}}{12},\frac{18-\sqrt{96\beta+36}}{12}]$, $\frac{6-\sqrt{96\beta+36}}6$, $d=\frac{M}{r}\frac{\iota^2}{r^2}$.
The homogenous solution \eqref{part} is not a physical one because it has no definite value as $r\to\infty$, and therefore we do not examine it further.
The particular solution of Eq. \eqref{V1} has the form
\be \omega^{(1,1)}_{Part.}=\frac{c_3}{r^3}\,,
\ee
where $c_3$ is a dimensional constant.

We obtain the non-vanishing metric components of the above solution
\begin{align}
\label{sol:metric_elements11}
g_{tt} =&\, -\left(\frac{r^2}{l^2}-\frac{M}{r}\right) - \frac {M}{r}\frac{a^2}{l^2}\, , \qquad \qquad g_{t \theta}= - \frac{M a} r  - \frac{c_1 \xi}{r}\equiv -\frac{\mathrm{const}}{r}\,, \nonumber \\
 \qquad
g_{rr} =&\,\frac{1}{\left(\frac{r^2}{l^2}-\frac{M}{l}\right)}\, , \qquad  \qquad
g_{\theta \theta} = \frac{r^2}{l^2}+ \frac{M}{r}\frac{a^2}{l^2}\,, \qquad g_{\rho\,\rho}=\frac{r^2}{l^2}\, .
\end{align}
Equation \eqref{sol:metric_elements11} shows that the  CS  scalar field  has no effect on the metric. This means that any spacetime asymptote as (A)dS of the CS scalar field will have no effect.
In the next section, we examine this conclusion and show that it is true if we include the charge field.

\section{Slowly charged rotating (A)dS BH solution}\label{charg}
In the charged BH solution, the Lagrangian of the gravitational theory is represented as
\be
\label{CSaction1}
L = L_{1} + L_{2} +  L_{3} +L_{4}+L_{5}\,,
\ee
where  $L_{1}$,  $L_{2}$,  $L_{3}$, $L_{4}$ are defined in Eqs.~ \eqref{EH-action}, \eqref{Theta-action}, \eqref{CS-action}, and \eqref{MM}, while $L_{E.M.}$ is the electromagnetic field, defined as
\be
S_4= \frac{\alpha_1 }{2} \int_{V}d^{4}x\sqrt{-g}\left(\frac{F^{2}}{2}\right)\,, \quad \textrm{where} \quad F^2=F_{\mu \nu}F^{\mu\nu}\,, \quad \textrm {and} \quad F_{\mu \nu} =2A_{[\mu, \nu]}\,,
\ee
where $A$ represents the gauge potential for electromagnetism.

The action's variation with respect to the metric and the electromagnetic gauge potential, as described in \eqref{CSaction1}, yields the following field equations:
\ba
\label{eom1}
&&R_{ab} + \frac{\alpha_1}{\kappa} C_{ab} = \frac{1}{2 \kappa} \left({\cal T}_{ab} - \frac{1}{2} g_{ab} {\cal T} \right),
\\
\label{eq:elct}
&&0=\partial_b \left( \sqrt{-g} {\textrm F}^{ab} \right)\,.
\ea
Conversely, when the action \eqref{CSaction1} is varied with respect to the CS term, it results in Eq.~ \eqref{eq:constraint}.

We employ the procedure outlined earlier to the modified   equations of motions and the  equation of evolution for the dynamical CS. Till ${\cal{O}(1,0)}$ we obtain the  evolution equation in the following format:
\ba
\label{1st-eq1}
&&{\mathit Q} \vartheta^{(1,0)}_{,rr} + \frac{1}{r} \vartheta^{(1,0)}_{,r} \left(\frac{4r^2}{l^2} - \frac{M}{r} \right)
=0 \,,
\ea
where ${\textit Q}=\left(\frac{r^2}{\iota^2} - \frac{M}{r}+\frac{q^2}{r^2} \right)$.
 Solution of the differential equation \eqref{1st-eq1} with homogeneous terms is represented as follows:
\be\label{theta-sol-SR111}
\vartheta^{(1,0)}(r) =-c_4\int\frac{r^2\,\iota^2}{\mathit Q}\, dr\approx \frac{c_4}{r^3}\left(1+\frac{M}{2r}\frac{\iota^2}{r^2}-\frac{3q^2}{7r^2}\frac{\iota^2}{r^2}\right)+{\cal O}\left(\frac{1}{r^9}\right)\,.
\ee
where $c_4$ is a dimensional constant of integration.

We deal with the second set $(t,\theta)$-component, which yields
\ba
\label{V11}
Q\omega^{(1,1)}_{rr}-\frac{4Q}{r}\omega^{(1,1)}_r-\frac{6\alpha_1}{\iota^2}\omega^{(1,1)} =6\alpha_1\left(\frac{M}{r}-\frac{q}{r^2}\right)\frac{a}{\iota}\frac{\alpha_1}{r\iota}\,. \ea
The equation above is a non-homogeneous differential equation with both homogeneous and particular solutions. Homogeneous solution is represented as follows:
 \ba\label{part1}
 &&\omega^{(1,1)}\approx c_5\,{r}^{-\frac{3+\,\sqrt {3}\sqrt {{{3\,+8\,\beta}}}}{2}}{e^{{\frac {\,\sqrt {6}{\iota}^{2}\alpha_1 \, \left( Mr-{q}^{2} \right) }{2\sqrt {-{q}^{2}{\iota}^{ 2}\alpha_1}{r}^{2}}}}}HeunB \Bigg[ \sqrt {3}\sqrt {{{3\, +8\,\alpha_1}}},{\frac {i\alpha_1\,{}^{2}M{\iota}^{2} \sqrt [4]{6}}{ \left( -{q}^{2}{\iota}^{2}\alpha_1 \right) ^{3/4}}},-{\frac {{\alpha_1}^{2}{M}^{2}{\iota}^{4} \sqrt {6}}{4 \left( -{q}^{2}{\iota}^{2}\alpha_1 \right) ^{3 /2}}},0,{\frac {i\sqrt [4]{6}\sqrt [4]{-{}^{3}{q}^{2}{\iota}^ {2}\alpha_1}}{\,r}} \Bigg] \nonumber\\
 &&+c_6\,{r}^{-\frac{3-\,\sqrt {3}\sqrt {{{3\,+8\,\beta}}}}{2}}HeunB \bigg[ - \sqrt {3}\sqrt {{ {3\,+8\,\alpha_1}}},{\frac {i\alpha_1\, {}^{2}M{\iota}^{2}\sqrt [4]{6}}{ \left( -{q}^{2}{\iota}^{2}\alpha_1 \right) ^{3/4}}},-{\frac {{\alpha_1}^{2}{M}^{2}{\iota}^{4}\sqrt {6}}{ 4\left( -{q}^{2}{\iota}^{2}\alpha_1 \right) ^{3/2}}},0,{\frac {i\sqrt [4]{6}\sqrt [4]{-{ }^{3}{q}^{2}{\iota}^{2}\alpha_1}}{\,r}} \Bigg] \nonumber\\
 &&\times\,e^{ - \frac{\left\{ i{r}^{2} \left( \sqrt {3}\sqrt {{\frac {3\,+8\, \alpha_1}{}}}-1 \right) \pi \,\sqrt {-{q}^{2}{\iota} ^{2}\alpha_1}-2\,\,\sqrt {6}{\iota}^{2}\alpha_1\, \left( Mr-{q}^{2} \right)  \right\}} {{4{r}^{2}\sqrt {-{q}^{2}{\iota}^{2} \alpha_1}}}}\,,
\ea
where $HeunB$ function is the solution of the Heun Biconfluent equation\footnote{The Heun function, $ HeunB(\alpha,\beta,\gamma,\delta,z)$, is the solution of the Heun Biconfluent equation which has the form \be Y''(z)-\frac{\beta z-\alpha+2z^2-1}{z}Y'(z)-\frac{2(\alpha-\gamma+2)z+\delta+\beta+\beta\alpha}{2z}Y(z)\,.\ee The above differential equation should satisfy the following boundary conditions: $Y(0)=1$, $Y'(0)=\frac{\delta+\beta+\beta\alpha}{2(1+\alpha)}$.}.
The homogenous solution of Eq. \eqref{V11} in Eq.\eqref{part1} is not a physical one because the term multiplied by the constant $c_6$ has no definite value as $r\to\infty$. Therefore, we set the constant $c_6=0$. Similarly, the term multiplied by the constant $c_5$ is a not real one, and hence we take the constant $c_5=0$. The particular solution of Eq. \eqref{V11} gives the form
\be
\omega^{(1,1)}_{Part.}=\frac{c_7}{r^3}\,,
\ee
where $c_3$ is a dimensional constant.

We obtain the non-vanishing metric components of the above solution
\begin{align}
\label{sol:metric_elements33}
g_{tt} =&\, -\left(\frac{r^2}{l^2}-\frac{M}{r}+\frac{q^2}{r^2}\right) - \left(\frac {M}{r}-\frac {q^2}{r^2}\right)\frac{a^2}{l^2}\, , \qquad \qquad g_{t \theta}= - \left(\frac{ M}{r}-\frac{q^2}{r^2}\right)\frac{a}\iota  - \frac{c_7 \xi}{r}\frac{a}\iota\equiv -\left(\frac{\mathrm{const}}{r}-\frac{q^2}{r^2}\right)\frac{a}\iota\,, \nonumber \\
 \qquad
g_{rr} =&\,\frac{1}{\left(\frac{r^2}{l^2}-\frac{M}{l}+\frac {q^2}{r^2}\right)}\, , \qquad  \qquad
g_{\theta \theta} = \frac{r^2}{l^2}+ \left(\frac{M}{r}-\frac{q^2}{r^2}\right)\frac{a^2}{l^2}\,, \qquad g_{\rho\,\rho}=\frac{r^2}{l^2}\, .
\end{align}
Equation \eqref{sol:metric_elements33} shows that the CS scalar field does not affect the metric in the charged case. This means that any spacetime asymptote as (A)dS in the CS scalar field will have no effect.

\section{Conclusion and discussions}
\label{conclusions}

In this article, we have investigated the (A)dS-Kerr spacetime by using the dynamical CS field equations. We have studied two cases in detail: i) The non-charged case and ii) the charged one.\\ In the non-charged case, it has been shown that the scalar field of the SC theory, Eq. \eqref{eq:constraint}, is affected compared with the solution presented in \cite{Yunes:2007ss}. Consequently, the components of the CS, i.e.,  $\vartheta^{(1,0)}$ differ from those presented in Ref.~\cite{Yunes:2007ss}. Despite this fact, if we use the scalar field $\vartheta^{(1,0)}$ in the   equation of motions \eqref{eom}, we do not obtain any extra terms in the metric, up to order $\epsilon$. Although the correction of $\omega^{(1,1)}$ of slow Kerr gives an asymptotic form of order ${\cal O}\left(\frac{1}{r^6}\right)$, we have shown that the asymptotic form of $\omega$ is not different from the form of $\omega^{(1,1)}$ up to  order $\xi$ in the case of (A)dS-Kerr spacetime.

In the charging case, we have followed the same procedure carried out for the case of the non-charged case. We  have shown  that the CS scalar field \eqref{eq:constraint} is affected compared to the slowly rotating Kerr solution discussed in Ref. \cite{Yunes:2007ss}. Moreover, we have shown that using this scalar field in the field equations \eqref{eom}, the asymptotic of $\omega^{(1,0)}$ has no effect on the asymptote of $\omega^{(1,1)}$ up to order $\xi^{1}$. It is ensured from the above two cases, charged and non-charged, that any rotating spacetime that asymptotes as (A)dS the cross-term, which is responsible for the rotation, cannot be affected by the CS scalar field, unlike the slowly rotating Kerr solution that asymptotes as flat spacetime.

Finally, we end this study by mentioning the following point. In this study, we have obtained a slow Kerr-(A)dS  BH solution in the dynamical  theory by employing a linear form of angular momentum, i.e.,   $a$,  without considering the potential. The issue that the addition of the potential may lead to any new physics different from (A)dS-Kerr without the potential will be studied in the near future.

\section*{Acknowledgments}

The work of KB was partially supported by the JSPS KAKENHI Grant
Number 21K03547.


\section*{Data Availability Statement}
No new data were created or analyzed in this study.

%
\end{document}